\documentclass[prl,twocolumn,twoside,preprintnumbers,superscriptaddress,nofootinbib]{revtex4}

\usepackage{epsfig}
\usepackage{graphicx}
\usepackage{latexsym}
\usepackage{bm}

\begin{document}

\title{Lepton-flavor universality limits in warped space}

\author{Eugenio Meg\'{\i}as}
\email{eugenio.megias@ehu.eus}
\affiliation{Departamento de F\'{\i}sica Te\'orica, Universidad del Pa\'{\i}s Vasco UPV/EHU, Apartado 644,  48080 Bilbao, Spain}

\author{Mariano Quir\'os}
\email{quiros@ifae.es}

\author{Lindber Salas}
\email{lsalas@ifae.es}

\affiliation{Institut de F\'{\i}sica d'Altes Energies (IFAE), and The Barcelona Institute of  Science and Technology (BIST)\\ Campus UAB, 08193 Bellaterra (Barcelona) Spain}

\date{\today}

\begin{abstract}
We explore the limits on lepton-flavor universality (LFU) violation in theories where the hierarchy problem is solved by means of a warped extra dimension. In those theories LFU violation, in fermion interaction with Kaluza-Klein modes of gauge bosons, is provided \textit{ab initio} when different flavor of fermions are differently localized along the extra dimension. As this fact arises from the mass pattern of quarks and leptons, LFU violation is natural in this class of theories. We analyze the experimental data pointing towards LFU violation, 
as well as the most relevant electroweak and flavor observables, and the LFU tests in the $\mu/e$ and $\tau/\mu$ sectors. We find agreement with $R_{K^{(\ast)}}$ and $R_{D^{(\ast)}}$ data at 95\% CL, provided the third generation left-handed fermions are composite ($0.14 < c_{b_L} < 0.28$ and $0.27 < c_{\tau_L} < 0.33$), and find the absolute limits  $R_{K^{(\ast)}}\gtrsim 0.79$ and $R_{D^{(\ast)}}/R_{D^{(\ast)}}^{\rm SM}\lesssim 1.13$. Moreover we predict $\mathcal B( B\to K\nu\bar\nu)\gtrsim 1.14\times 10^{-5}$ at 95\% CL, smaller than the present experimental upper bound but a few times larger than the Standard Model prediction.
\end{abstract}



\maketitle



\textit{I. Introduction.--} In recent years lepton-flavor universality (LFU) has been challenged by several experimental results. The LHCb Collaboration has determined the branching ratio $\mathcal B(\bar B\to\bar K\ell\ell)$ ($\ell=\mu,e$) for muons over electrons, for $1<q^2/GeV^2<6$, yielding~\cite{Aaij:2014ora}
\begin{equation}
R_K=\frac{\mathcal B(\bar B\to\bar K\mu\mu)}{\mathcal B(\bar B\to\bar K ee)}=0.745^{+0.090}_{-0.074}\pm 0.032 \,,
\label{RK}
\end{equation}
and very recently the same tendency has been shown for the ratio~\cite{Aaij:2017vbb}
\begin{equation}
R_{K^\ast}=\frac{\mathcal B(\bar B\to\bar K^\ast\mu\mu)}{\mathcal B(\bar B\to\bar K^\ast ee)}=\left\{
\begin{array}{c} 0.660 ^{+0.110}_{-0.070}\pm 0.024\\
0.685^{+0.113}_{-0.069}\pm 0.047
\end{array}\right.
\end{equation}
 for $0.045<q^2/GeV^2<1.1$ and $1.1<q^2/GeV^2<6$, respectively. Given that the Standard Model (SM) prediction is $R_{K^{(\ast)}}=1.00\pm 0.01$~\cite{Bordone:2016gaq}, the departure of both observables is $\sim 2.5\sigma$, which suggests LFU violation in the process $b\to s\ell\ell$.
 
 Moreover, the charged current decays $\bar B\to D^{(\ast)} \ell^-\bar\nu_\ell$ have been measured by the BaBar~\cite{Babar:ref}, Belle~\cite{Belle:ref} and LHCb~\cite{Aaij:2015yra} Collaborations which measure
\begin{equation}
R_{D^{(\ast)}}=\frac{\mathcal B(\bar B\to D^{(\ast)}\tau^-\bar\nu_\tau)}{\mathcal B(\bar B\to D^{(\ast)}\ell^-\bar\nu_\ell)},\quad (\ell=\mu \textrm{ or } e) \ .
\end{equation}
The averaged experimental results 
\begin{equation}
R_D=0.403\pm 0.047,\quad R_{D^\ast}=0.310\pm 0.017
\end{equation}
again depart from the SM predictions 
$R_D^{\rm SM}=0.300\pm 0.011$, $R_{D^\ast}^{\rm SM}=0.254\pm 0.004$
by $\sim 2.2\sigma$ and $3.3\sigma$, respectively, although the combined deviation is $\sim 4\sigma$. This again suggests LFU violation in the process 
$b\to c\tau\bar\nu_\tau$.

If (some of) the previous $B$-anomalies were confirmed, they would constitute an inevitable proof of the existence of new physics (NP) beyond the SM. In this letter we would concentrate in a class of NP scenarios, motivated by the solution of the SM hierarchy problem, where LFU violation is implemented \textit{ab initio}: models with a warped extra dimension~\cite{Randall:1999ee}, where SM fermions are propagating in the bulk. 

Theories with a warped geometry are characterized by a five-dimensional (5D) metric $A(y)$, such that $ds^2=e^{-2A}\eta_{\mu\nu}dx^\mu dx^\nu+dy^2$, and two branes located at the ultraviolet (UV)  $y=0$, and infrared (IR) $y=y_1$, such that $A(y_1)\simeq 35$ to solve the SM hierarchy problem. In the absence of a stabilizing scalar field $\phi$ the metric is of type $AdS_5$~\cite{Randall:1999ee} and the radion potential is flat. To fix the interval length we have to introduce the stabilizing field $\phi$ with a bulk potential, and brane potentials fixing its values at $\phi(0)$ and $\phi(y_1)$. A simple way of solving the 5D gravitational equations of motion is by introducing a superpotential function $W(\phi)$, which transforms the original equations into linear differential equations that can easily be solved. 
%
%
%
In this paper we will consider the case where $W(\phi)$ goes asymptotically as an exponential~\cite{ref:model}, and the metric acquires a strong deformation of conformality near the IR brane, which triggers a small mismatch between the physical Higgs profile and the gauge KK modes, thus suppressing the electroweak observables. 

Moreover we will consider the superpotential $W(\phi)=6k(1+e^{a\phi})^{b}$~\cite{ref:model} where $a$ and $b$ are real parameters. We will make the choice $a=0.15$, $b=2$, as in Ref.~\cite{Megias:2017ove} where many details of the formalism can be found.
In this case we can consider all the SM fields as 5D fields propagating in the bulk of the extra dimension. In particular the SM fermions $f_{L,R}$ are the zero modes of 5D fermions $\Psi(y,x)$ with appropriate boundary conditions and a 5D Dirac mass term
$M_{f_{L,R}}(y)=\mp c_{f_{L,R}}W(\phi(y))$. Partly composite fermions in the dual holographic theory, with $c_{f_{L,R}}<0.5$ (almost elementary fermions, with $c_{f_{L,R}}>0.5$) are localized towards the IR (UV) brane and thus interact strongly (weakly) with KK modes, which are IR localized. As we will see, assuming a different degree of compositeness for different lepton and quark flavors in the dual theory amounts to imposing the LFU violation that experimental data seem to suggest. 

\textit{II. The model.--} We will now consider a 4D theory which contains, on top of the SM zero mode gauge, fermion and Higgs fields, the KK excitations of gauge bosons $Z^n_\mu$, $\gamma^n_\mu$, $W^n_\mu$. Interaction between zero modes of fermion, Higgs and gauge boson fields are identical to the SM ones. Interaction of neutral gauge KK modes $X^n=Z^n,\gamma^n$, with profiles $f_X^n(y)$, with zero mode fermions, with profiles $f_{L,R}(y)$ in the weak basis, is given by the neutral current Lagrangian
\begin{equation}
\mathcal L=\frac{g}{c_W}\sum_{X,n} X_n^\mu\left(\bar f_L g_{f_L}^{X_n}\gamma_\mu f_L+\bar f_R g_{f_R}^{X_n}\gamma_\mu f_R \right) \,,
\label{neutrallagrangian}
\end{equation}
where $g^{X_n}_{f_{L,R}}=g^{X}_{f_{L,R}} G_{f_{L,R}}^{\,n}$, and $X=Z,\gamma$, with the SM couplings
%
$g_{f}^Z=(T_{3f}-Q_f s^2_W),\,   g_{f}^\gamma=Q_f s_W c_W$,
%
and the overlapping integrals $G^n_{f}$ defined as
\begin{equation}
G^n_{f}=\frac{{\displaystyle \sqrt{y_1} \int e^{-3A} f_X^n(y) f^2(y)}}{\sqrt{\int [f_X^n(y)]^2 } \int e^{-3A}  f^2(y)  }  \, .
\label{integral}
\end{equation}
Similarly the charged current interaction of KK modes $W^n_\mu$ with zero mode fermions
is given by the Lagrangian
\begin{equation}
\mathcal L=\frac{g}{\sqrt{2}}\sum_n W^n_\mu\left[\bar u_i G_{d_L^i} P_L \gamma^\mu d_i+ \bar\ell_i \gamma^\mu G_{\ell_L^i}P_L\nu_i  \right] \,.
\label{chargedlagrangian}
\end{equation}

After electroweak symmetry breaking the quark mass matrix is diagonalized by unitary transformations $V_{u_{L,R}}$ and $V_{d_{L,R}}$. Flavor-changing neutral, and charged, currents are generated in the mass eigenstate basis from Lagrangians (\ref{neutrallagrangian}) and (\ref{chargedlagrangian}) by 
$\left(V^\dagger_{d_{L,R}}G_{d_{L,R}}^nV_{d_{L,R}}\right)_{ij}$ and $\left(V^\dagger_{u_L}G_{d_ L}^nV_{d_L}\right)_{ij}$, as the matrices $G_{d_{L,R}}^n=\textrm{diag}(G^n_{d_{L,R}},G^n_{s_{L,R}},G^n_{b_{L,R}})$ are diagonal but not proportional to the unit matrix. This phenomenon generates LFU violation, absent in the SM, which can explain experimental data on $B$-anomalies. In particular $b\to s$, and $b\to c$, transitions are generated by the previous elements with ($i=3, j=2$), i.e.~$(V^\ast_{d_{L,R}})_{k3}G_{d_{L,R}^k}^n(V_{d_{L,R}})_{k2}$ and $(V^\ast_{u_{L}})_{k3}G_{d_{L}^k}^n(V_{d_{L}})_{k2}$, respectively.
In the lepton sector we will neglect neutrino masses. Moreover to prevent \textit{lepton flavor violation} in our theory, we are assuming that the 5D Yukawa couplings are such that the charged leptons are diagonal in the weak basis, so that $V_{\ell_{L,R}}\simeq 1$. 

In the absence of a general (UV) theory, providing the 5D Yukawa couplings, we will just consider the general form for these matrices by assuming that they reproduce the physical CKM matrix $V$, i.e.~they satisfy the condition $V\equiv V_{u_L}^\dagger V_{d_L}$. Given the hierarchical structure of the quark mass spectrum and mixing angles, we can then assume for the matrices $V$, $V_{d_L}$ and $V_{u_L}$ Wolfenstein-like parametrizations as 
\begin{equation}
V_{d_L}=\left(\begin{array}{ccc} 1-\frac{1}{2}\lambda_0^2 &\lambda_0&  (V_{d_L})_{13}\\
-\lambda_0 & 1-\frac{1}{2}\lambda_0^2 & A \lambda^2(1-r)\\
(V_{d_L})_{31} &\quad -A \lambda^2(1-r) &1
\end{array}\right)
\label{Vd}
\end{equation}
where $(V_{d_L})_{13}=A \lambda^2\lambda_0(1-r)(\rho_0-i \eta_0)$ and $(V_{d_L})_{31}=A\lambda^2\lambda_0(1-r)(1-\rho_0-i\eta_0)$, with $(\lambda,\rho,\eta)$ the parameters of~$V$. The matrix $V_{u_L}$ is obtained from the condition $ V_{u_L}= V_{d_L}V^\dagger$. The values of $(r,\lambda_0,\rho_0,\eta_0)$ should be consistent with the hierarchical structure of the matrix. Moreover we will assume that $V_{u_R}\simeq V_{u_L}$ and $V_{d_R}\simeq V_{d_L}$.

In the following we will also assume that the first and second generation quarks respect the universality condition. This implies an approximate accidental
$U(2)_{q_L} \otimes U(2)_{u_R} \otimes U(2)_{d_R}$ global flavor symmetry, which is only broken by the Yukawa couplings~\cite{Megias:2016bde}. 
For simplicity in our numerical analysis we will moreover choose $c_{q^{1}_L}=c_{q^{2}_L}\equiv c_{q_L} = 0.60$, as well as $c_{u_R}=c_{c_R}=c_{d_R}=c_{s_R}\equiv c_{q_R} = 0.60$.

Finally the leading flavor-violating couplings of the KK gluons $G_{n\mu}^A$ involving the down quarks are given by
\begin{eqnarray}
\mathcal L_s^d
=& g_sG_{n\mu}^A\Big[
\,\bar d_i\gamma^\mu t_A\left\{ (V_{d_L}^*)_{3i} (V_{d_L})_{3j}\left(G_{b_{L}}^n-G_{q_{L}}^n  \right)P_L\right.\nonumber\\
+&\left.(V_{d_R}^*)_{3i} (V_{d_R})_{3j}\left(G_{b_{R}}^n-G_{q_{R}}^n  \right)P_R\right\} d_j + {\rm h.c.} \Big]\label{eq:lagrangian_QCD}
\end{eqnarray}
where $t_A$ are the $SU(3)$ generators in the triplet representation, and a similar expression holds for the up sector $\mathcal L_s^u$ by replacing $d \to u$ and $b \to t$ in (\ref{eq:lagrangian_QCD}).

\textit{III. The effective theory.--} After integrating out the gauge boson KK modes $Z^n$ and $\gamma^n$ we obtain the effective operators contributing to the $\Delta F=1$ transitions 
\begin{equation}
\mathcal L_{eff}=\frac{G_F\alpha}{\sqrt{2}\pi}V^\ast_{ts}V_{tb}\sum_i C_i\mathcal O_i  \,,
\end{equation}
where the Wilson coefficients $C_i=C_i^{\rm SM}+\Delta C_i$ are the sum of the SM contribution, $C_i^{\rm SM}$, and the NP one, $\Delta C_i$. The sum includes the semileptonic operators
\begin{equation}
\mathcal O_{9}^{(\prime)\ell} =(\bar s\gamma_{\mu} P_{L(R)} b)(\bar\ell \gamma^\mu\ell),\
\mathcal O_{10}^{(\prime)\ell} =(\bar s\gamma_{\mu} P_{L(R)} b)(\bar\ell \gamma^\mu\gamma_5\ell)
\end{equation}
and the corresponding Wilson coefficients are given by
\begin{equation}
\displaystyle \Delta C_{9,10}^{(\prime)\ell}=\mp(1-r)\sum_{X,n}\frac{2\pi g^2 g_{\ell_{V,A}}^{X_n}\left(g^{X_n}_{b_{L(R)}}-g^{X_n}_{q_{L(R)}}\right)}{\sqrt{2} G_F\alpha c_W^2 M^2_n}
\label{C9}
\end{equation}
where $g^{X_n}_{f_{V,A}}=(g^{X_n}_{f_{L}}\pm g^{X_n}_{f_{R}})/2$, while $C_9^{\rm SM}\simeq -C_{10}^{\rm SM}\simeq 4.2$. In particular the $R_K$ observable in Eq.~(\ref{RK}) is given by
\begin{equation}
R_K=\frac{|C_9^\mu+C_9^{\prime\mu}|^2+|C_{10}^\mu+C_{10}^{\prime\mu}|^2}{|C_9^e+C_9^{\prime e}|^2+|C_{10}^e+C_{10}^{\prime e}|^2} \,.
\end{equation}

We also obtain the effective Lagrangian 
\begin{equation}
\mathcal L_{eff}=\frac{C^{t\ell}_n}{M_n^2}  (\bar t_L\gamma_\mu t)(\ell_L \gamma^\mu \ell_L)+\frac{C^{b\nu}_n}{M_n^2}(\bar t_L\gamma_\mu b_L)(\bar\tau \gamma_\mu \nu)
\label{tops}
\end{equation}
where
\begin{equation}
C^{t\ell}_n = -\frac{g^2}{c_W^2}\left( g_{u_L}^{Z_n} g_{\ell_L}^{Z_n}+g_{u_L}^{\gamma_n} g_{\ell_L}^{\gamma_n} \right),\, C^{b\nu}_n=-\frac{g^2}{2}G^n_{b_L}G^n_{\tau_L}.
\label{WC}
\end{equation}
The Lagrangian (\ref{tops}) will give rise after radiative corrections to leading modifications of the coupling, $\Delta g_{\ell_L}^Z$, and the LFU violation process $\tau\to\ell\nu\nu$, proportional to $Y_t^2$~\cite{Feruglio:2016gvd}, where $Y_t$ is the top Yukawa coupling, as it has been seen in Refs.~\cite{Feruglio:2016gvd,Megias:2017ove}.

Similarly, after integrating out the KK modes $W^n$ we obtain the effective operator
\begin{equation}
\mathcal L_{eff}=-\frac{4G_F}{\sqrt{2}} V_{cb} \sum_\ell C^{\ell}(\bar c\gamma^\mu P_L b)(\bar\ell\gamma_\mu \nu_\ell)
\label{effectiveL}
\end{equation}
with Wilson coefficients
\begin{equation}
C^{\ell} =\sum_n (m_W^2/m_{W^{(n)}}^2)\left[ G_{q_L}^n+ r(G_{b_L}^n-G_{q_L}^n)
\right]G_{\ell_L}^n \,.
\label{Csimp}
\end{equation}
The corrections to the $R_{D^{(\ast)}}$ observables from the effective operators are given, in terms of the Wilson coefficients, as~\cite{Bhattacharya:2016mcc}
\begin{equation}
R_{D^{(\ast)}}(C^\tau,C^\mu)=2 R_{D^{(\ast)}}^{\rm SM}\frac{\left|1+
C^\tau  \right|^2
}{1+\left|1+
C^\mu  \right|^2 
}\ .
\label{expresion}
\end{equation}

Finally NP contributions to $\Delta F=2$  processes  come  from  the  exchange of gluon  KK modes $G_n$. After integrating out $G_n$, the couplings in Eq.~(\ref{eq:lagrangian_QCD}) give rise to a set of $\Delta F=2$ dimension-six operators~\cite{Megias:2016bde} as e.g.
\begin{equation}
{\cal L}_{\Delta F = 2} = \sum_n \frac{c_{dij}^{LL(n)}}{M_n^2} (\overline d_{iL} \gamma^\mu d_{jL}) (\overline d_{iL} \gamma_\mu d_{jL})
\label{flavor1}
\end{equation}
where
\begin{equation}
c^{LL(n)}_{dij}  = \frac{g_s^2}{6}\left[(V_{d_L}^*)_{3i} (V_{d_L})_{3j}\right]^2\left(G_{b_{L}}^n-G_{q_{L}}^n  \right)^2 \,,
\label{flavor2}
\end{equation}
and similarly for other chiralities and for up quarks. For more details see Ref.~\cite{Megias:2016bde}.

\textit{IV. Fitting the B-anomalies.--} We will now adjust the $B$-anomalies in the $b\to s(c)\ell\ell$ processes. We have to first fix the model parameter $r$ in the transformation matrices. In Ref.~\cite{Megias:2017ove} we have used $r<1$ (and composite $\mu_L$) to avoid any possible fine-tuning ($\sim 100/r$ \%) in the determination of the CKM matrix out of the transformations $V_{u_L,d_L}$. However as the observables $R_{D^{(\ast)}}$ are parametrically $\propto r$ we found a strong tension with electroweak observables. Here we will relax the previous condition and explore values $r>1$ for which, as we will see, the solution to the $R_{K^{(\ast)}}$ anomaly is qualitatively different, as $\mu_L$ turns out to be elementary. We will consider for the moment $r=2.3$ (and $m_{KK}=2$ TeV) for which the fine-tuning is pretty moderate ($\sim 40\%$).
\begin{figure}[htb]
\centering
\includegraphics[width=7.4cm]{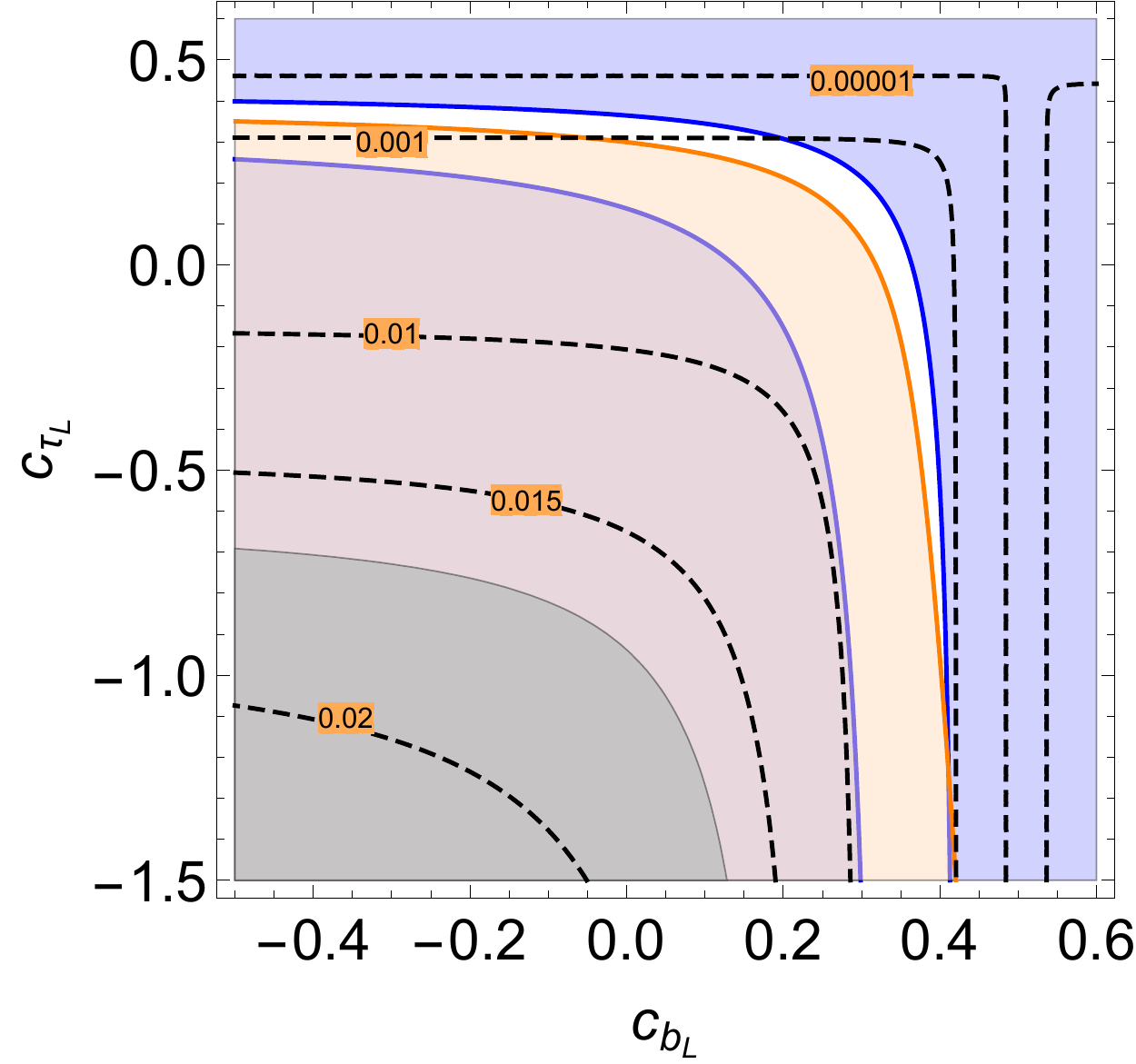} 
\caption{\it Allowed region coming from $R_{D^{(\ast)}}$ (white region) and forbidden from $\sigma(b\bar b\to Z^n)\mathcal B(Z^n\to \tau\tau)$ in pb (grey region). The orange region is forbidden by the constraint on $g_{\tau_L}^Z$.
%
}
\label{fig:RDsigmaBr}
\end{figure} 
 In Fig.~\ref{fig:RDsigmaBr} we show in the plane $(c_{b_L},c_{\tau_L})$ the allowed region at 95\% CL  (between the blue solid lines) from the experimental values of $R_{D{(\ast)}}$ (blue region is excluded). We superimpose the experimentally excluded region from direct $\tau\tau$ production in bottom-bottom fusion~\cite{bbfusion:ref} (grey band) and the contour lines for $\sigma(b\bar b\to Z^n)\mathcal B(Z^n\to \tau\tau)$ in pb which can provide in the future stronger constraints. We have considered $c_{\mu_L} = 0.60$ and $c_{t_R} = 0.50$.
\begin{figure}[htb]
\centering
\includegraphics[width=7.4cm]{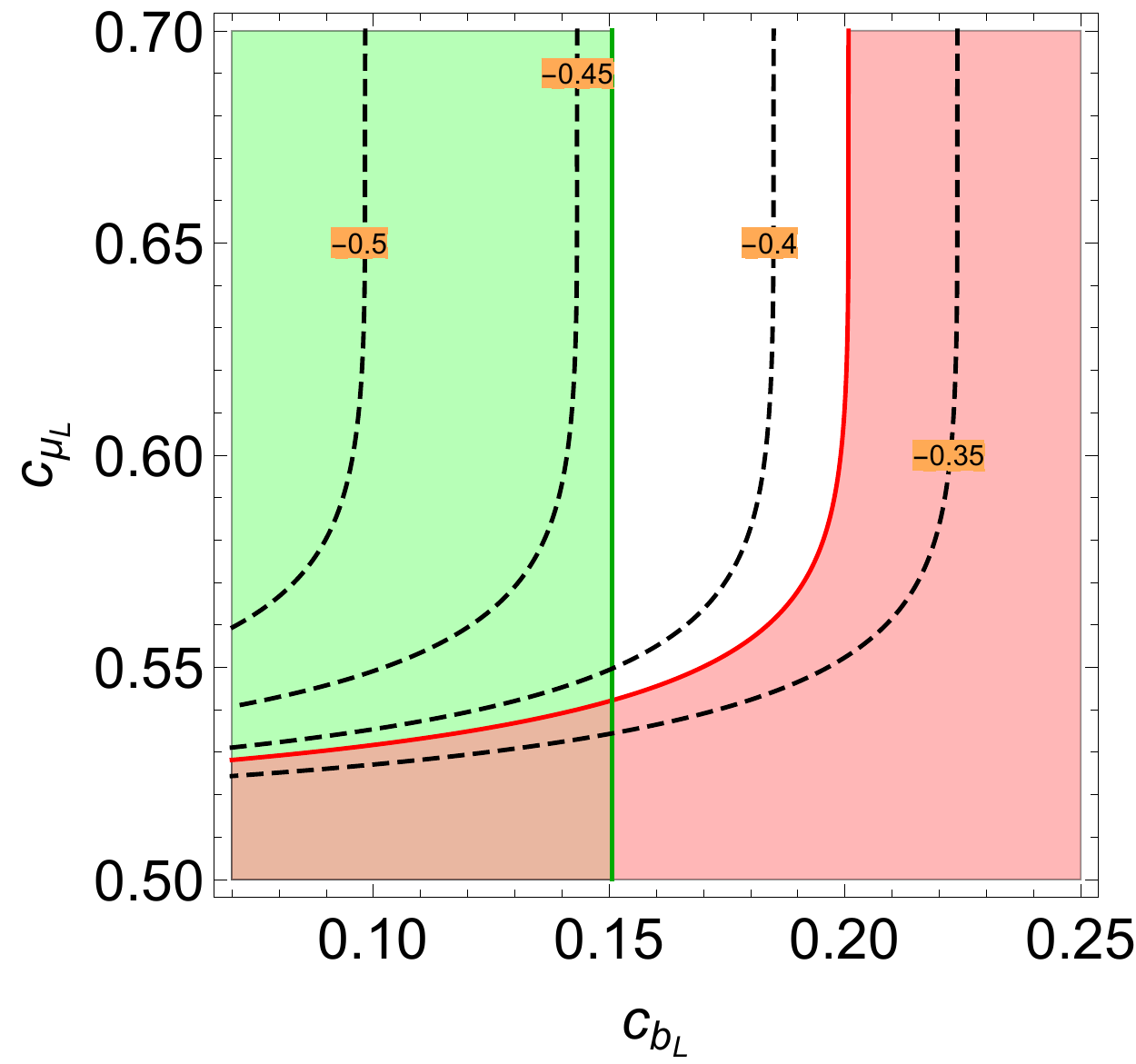} \hfill
\caption{\it 
Region in the $(c_{b_L},c_{\mu_L})$ plane that accommodates the $2\sigma$ region $\Delta C_9^\mu$. The green band corresponds to the excluded region by flavor constraints.
}
\label{fig:C9}
\end{figure} 
Global fits to $\Delta C_{9,10}^{(\prime)\,\mu}$ have also been performed in the literature by using a set of observables, including the branching ratios for $B\to K^* \ell\ell$, $B_s\to\phi\mu\mu$ and $B_s\to\mu\mu$, in Refs.~\cite{Globalfit:ref1,Globalfit:ref2,Altmannshofer:2017fio}. As experimental data favor $\Delta C_9^\mu\simeq -\Delta C_{10}^\mu$, we have fixed $c_{\mu_R}=0.5$ which provides the latter relation. We have also used $c_{e_L}=0.5$ which leads to $\Delta C_9^e\simeq \Delta C_{10}^e$, consistent with the $R_{K^{(\ast)}}$ anomaly. In addition we have considered $c_{e_R}=0.85$ and $c_{b_R}=0.55$. Using the recent multi-observable fit (which includes $R_{K^{(\ast)}}$) from Refs.~\cite{Globalfit:ref1,Altmannshofer:2017fio} we get the $2\sigma$ interval $\Delta C_9^\mu\in [-0.99,-0.38]$. We show in the plot of Fig.~\ref{fig:C9} the region in the $(c_{b_L},c_{\mu_L})$ plane that accommodates the previous constraint on $\Delta C_9^\mu$, which leads to $c_{b_L} \lesssim 0.20$, along with contour lines of $\Delta C_9^\mu$.

\textit{V. Constraints.--} 
The main constraints are those from the experimental value of the coupling $g_{\tau_L}^Z$ and LFU tests, as e.g.~$\tau\to\mu\nu\nu$, as well as constraints from flavor physics. 
 The SM value of $g_{\tau_L}^Z$ receives tree-level corrections from KK modes of gauge bosons and fermions and leading loop corrections to the Wilson coefficient $C_n^{t\tau}$ in (\ref{WC}) proportional to $Y_t^2$
\begin{equation}
\Delta g_{\ell_L}^Z\simeq\frac{v^2}{M_n^2}\frac{1}{16\pi^2}\left(3Y_t^2C^{t\ell}_n \log\frac{M_n}{m_t} +\mathcal O(g^4)\right)\,.
\label{eq:Delta_g}
\end{equation}
Using the experimental value from the fit of Ref.~\cite{ALEPH:2005ab} 
$g_{\tau_L}^Z=-0.26930\pm 0.00058$, we can exclude a region in the plane $(c_{b_L},c_{\tau_L})$, the orange region in Fig.~\ref{fig:RDsigmaBr}.

\begin{figure}[htb]
\centering
\includegraphics[width=7.4cm]{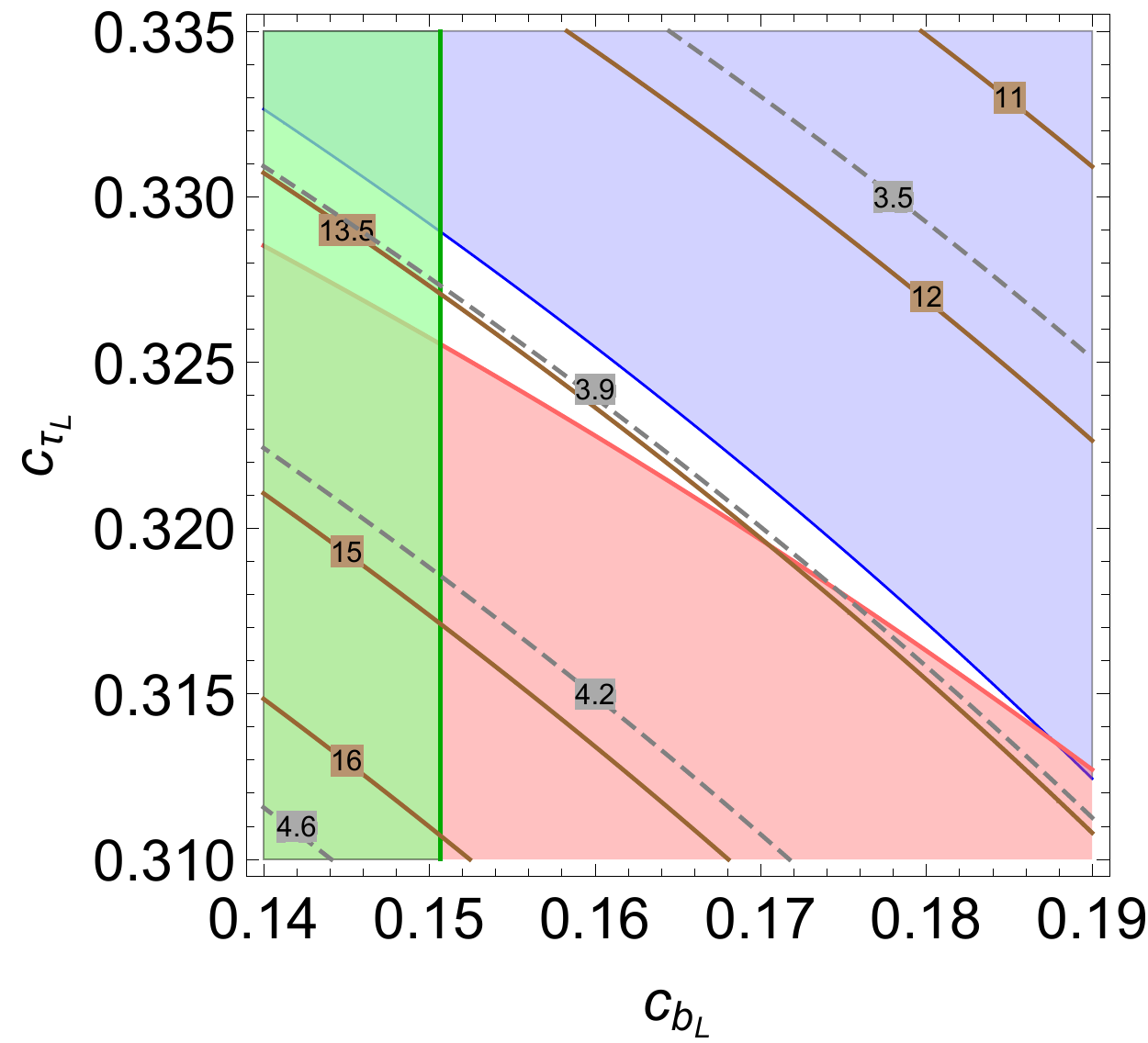} 
\caption{\it Red region in the plane $(c_{b_L},c_{\tau_L})$ is forbidden by $R_\tau^{\,\tau/\mu}$, while blue region is forbidden by $R_{D^{(\ast)}}$. Dashed and solid (brown) lines are contour lines for $R_{K^{(\ast)}}^\nu$ and $R_{K^{(\ast)}}^\tau$, respectively. The green region is forbidden by flavor constraints. We have considered $r=2.3$.}
\label{fig:RKtau}
\end{figure}

The value of $R_{D^{(\ast)}}$ has also to agree with flavor universality tests in tau decays. In particular the observables
\begin{equation}
R_\tau^{\tau/\ell}=\frac{\mathcal B(\tau\to\ell\nu\bar\nu)/\mathcal B(\tau\to\ell\nu\bar\nu)_{\rm SM}}{\mathcal B(\mu\to e\nu\bar\nu)/\mathcal B(\mu\to e\nu\bar\nu)_{\rm SM}},\quad (\ell=\mu,e)
\end{equation}  
are subject to the experimental bounds~\cite{Pich:2013lsa,Feruglio:2016gvd}, $R_\tau^{\tau/\mu}\in [0.996,1.008]$ and $R_\tau^{\tau/e}\in [1.000,1.012]$ at 95\% CL. In our model, fixing $c_{e_L}=0.5$ implies that $R_\tau^{\tau/e}=1$ while, including the relevant one-loop radiative corrections to $C_n^{b\nu}$~\cite{Feruglio:2016gvd}, we can write the $R_\tau^{\tau/\mu}$ observable as~\cite{Feruglio:2016gvd,Megias:2017ove}
\begin{equation}
R_\tau^{\tau/\mu} = 1 + 2 \frac{m_W^2}{m^2_{W^{(n)}}} G^n_{\tau_L} (G^n_{\mu_L} - 0.065 G^n_{b_L})  \,.  \label{eq:Rtau1loop}
\end{equation}
The region excluded in the $(c_{b_L},c_{\tau_L})$ plane is shown in red in Fig.~\ref{fig:RKtau} which is a blow up of Fig.~\ref{fig:RDsigmaBr}. The allowed region by both $R_\tau^{\tau/\mu} $ and $R_{D^{(\ast)}}$ is the white region in Fig.~\ref{fig:RKtau}. We have considered $c_{\mu_L}=0.60$.

Finally flavor constraints have been thoroughly considered in Refs.~\cite{Megias:2016bde,Megias:2017ove}. In this paper we will consider the leading (model independent) constraints coming from LL chiralities in Eqs.~(\ref{flavor1}) and (\ref{flavor2}). Indeed constraints coming from RR chiralities are not important, as we are considering the third generation right-handed quarks UV localized, while constraints coming from LR chiralities are always subleading for appropriate regions of the 5D Yukawa couplings $\sqrt{k}Y^{\rm 5D}_t$, as we have seen in Ref.~\cite{Megias:2017ove}.
In the present case (i.e. for $r=2.3$ and $m_{KK}=2$ TeV), and considering $\lambda_0 \simeq \lambda$, satisfying strong flavor constraints implies small values for the parameters $\eta_0$ and $\rho_0$ in (\ref{Vd}). In particular they must satisfy the bounds $\eta_0\lesssim 0.01\,\eta$ and $\rho_0\lesssim 0.2\, \rho$. Moreover we find the lower bound $c_{b_L}\gtrsim 0.15$, the green band in Figs.~\ref{fig:C9} and~\ref{fig:RKtau}. 

\textit{VI. Predictions.--} If there is a contribution to the process $B\to K^{(\ast)} \mu\mu$, contributions to $B\to K^{(\ast)} \bar\nu\nu$ and $B\to K^{(\ast)} \tau\tau$ will also be generated and will constitute the smoking guns for this model.
For the process $B\to K^{(\ast)} \nu\bar\nu$ we define the observable $R_{K^{(\ast)}}^\nu$, as
\begin{equation}
\frac{\mathcal B(B\to K^{(\ast)} \nu\bar\nu)}{\mathcal B(B\to K^{(\ast)} \nu\bar\nu)_{\rm SM}}\simeq\frac{\sum_\ell |C_\nu^{\rm SM}+(\Delta C_{\nu_\ell}+\Delta C_{\nu_\ell}^\prime)|^2}{3|C_\nu^{\rm SM}|^2} 
\end{equation}
where $\mathcal B(B\to K \nu\bar\nu)_{\rm SM}=(3.98\pm 0.47)\times 10^{-6}$, $\mathcal B(B\to K^{\ast} \nu\bar\nu)_{\rm SM}=(9.18\pm 1.00)\times 10^{-6}$~\cite{Buras:2014fpa}, and $C_{\nu}^{\rm SM}=-6.4$.  The $R_{K^{(\ast)}}^{\,\nu} $ observables are encoded by the operators
\begin{equation}
\mathcal O^{(\prime)}_ {\nu_\ell}=(\bar s_{L(R)}\gamma^\mu b_{L(R)})(\bar \nu_\ell  \gamma_\mu(1-\gamma_5)\nu_\ell)  
\end{equation}
and Wilson coefficients, 
\begin{equation}
\Delta C_{\nu_\ell}^{(\prime)}=-\frac{(1-r)\pi g^2 (g^{Z_n}_{b_{L(R)}}-g^{Z_n}_{s_{L(R)}})g^{Z_n}_{\nu_{\ell}}}{\sqrt{2}G_F\alpha c_W^2 M_n^2} \,.
\end{equation}
The experimentally excluded region at 90\% CL is~\cite{Lees:2013kla} $\mathcal B(B\to  K^{(\ast)} \nu\bar\nu)>1.6\times 10^{-5}\,(2.7\times 10^{-5})$.
In Fig.~\ref{fig:RKtau} we show the (dashed) contour lines of $R_{K^{(\ast)}}^{\,\nu}$ for $r=2.3$. The allowed 
 region is consistent with $3.85\lesssim R_{K^{(\ast)}}^\nu\lesssim 3.96$, leading to 
$1.18\times 10^{-5}\lesssim \mathcal B(B\to  K \nu\bar\nu)\lesssim 1.94\times 10^{-5}$, $2.78\times 10^{-5}\lesssim \mathcal B(B\to  K^\ast \nu\bar\nu)\lesssim4.41\times 10^{-5}$ at $95\%$ CL.

A similar analysis can be done with the ratio $R_K^\tau$, as
\begin{equation}
\frac{\mathcal B( B\to K \tau\tau)}{\mathcal B( B\to K \tau\tau)_{\rm SM}}=\frac{\left|C_9^{\tau}+C_9^{\prime\,\tau} \right|^2+\left|C_{10}^{\tau}+C_{10}^{\prime\,\tau} \right|^2}{2\left|C_9^{\rm SM} \right|^2} 
\end{equation}
which has been measured by the BaBar Collaboration providing the 90\% CL bound $\mathcal B( B\to K\tau\tau)< 2.25 \times 10^{-3}$~\cite{TheBaBar:2016xwe}, much larger than the SM prediction $\mathcal B(B\to K\tau\tau)_{\rm SM}=(1.44\pm 0.15)\times 10^{-7}$~\cite{Bouchard:2013mia}. We also show in Fig.~\ref{fig:RKtau} (solid) contour lines of constant $R_K^\tau$, where we have used $c_{\tau_R}=0.55$. From this plot we see that the allowed region is consistent with $13.2\lesssim R_K^\tau\lesssim 13.7$, pretty far from forthcoming experimental results.

\begin{figure}[htb]
\centering
\includegraphics[width=7.4cm]{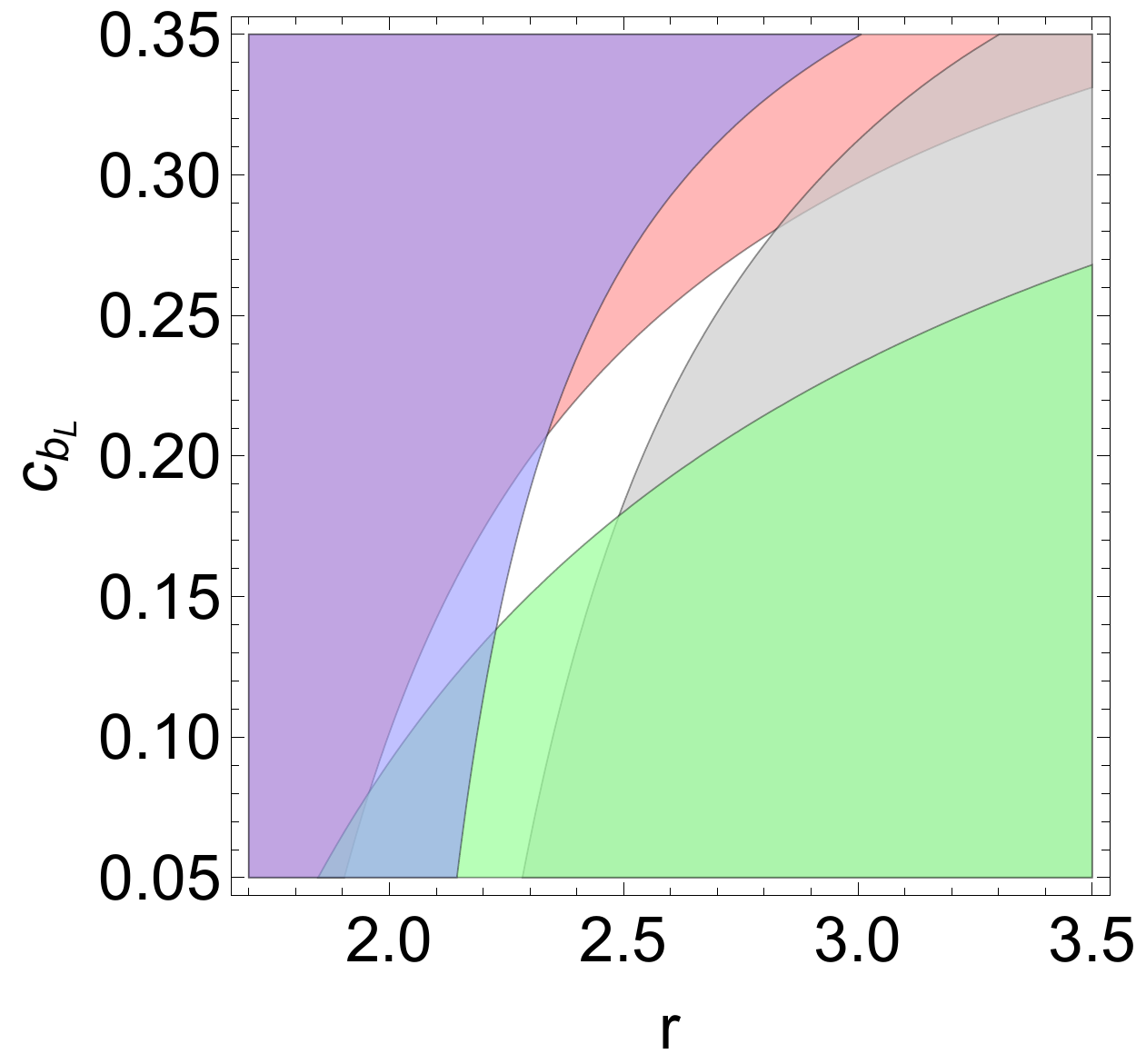} \hfill
\caption{\it Region in the space of parameters $(r,c_{b_L})$ compatible with $\Delta C_9$ (red), $R_{D^{(\ast)}}$, $R_\tau^{\tau/\mu}$ (blue), and flavor constraints: down quarks (grey) and up quarks (green). We have considered $c_{\mu_L}=0.6$.
}
\label{fig:flavour}
\end{figure}

\textit{VII. Conclusions.--} In this paper we have explored the capabilities of warped models to accommodate LFU violation. Throughout we have used the particular value $r=(V_{u_L})_{32}/V_{cb}=2.3$, which amounts to a 40\% tuning in the determination of the CKM matrix. The range of possible values of $r$ consistent with all experimental data is shown in Fig.~\ref{fig:flavour} from where we obtain the range $2.2 \lesssim r \lesssim 2.8$.
We find agreement with $R_{K^{(\ast)}}$ and $R_{D^{(\ast)}}$ data at 95\% CL, provided the third generation of left-handed fermions is composite, as $0.14 < c_{b_L} < 0.28$ and $0.265 < c_{\tau_L} < 0.33$. Moreover we obtain the absolute limits $R_{K^{(\ast)}} > 0.79$ and $R_{D^{(\ast)}}/R_{D^{(\ast)}}^{\rm SM} < 1.13$. Finally our model predicts, for any value of the parameters, the absolute range $3.74\lesssim R_{K^{(\ast)}}^\nu\lesssim 5.20$ , leading to 
$1.14\times 10^{-5}\lesssim \mathcal B(B\to  K \nu\bar\nu)\lesssim 2.55\times 10^{-5}$, $2.70\times 10^{-5}\lesssim \mathcal B(B\to  K^\ast \nu\bar\nu)\lesssim 5.79\times 10^{-5}$ at $95\%$ CL,
on the verge of experimental discovery or exclusion.

\vspace{0.2cm}
\begin{acknowledgments}
{\bf Acknowledgments}
We thank A. Ishikawa for pointing out to us the Belle Collaboration papers in Ref.~\cite{Lees:2013kla}. LS is supported by a \textit{Beca Predoctoral Severo Ochoa} of Spanish MINEICO (SVP-2014-068850), and EM is supported by the \textit{Universidad del Pa\'{\i}s Vasco} UPV/EHU, Bilbao, Spain, as a Visiting Professor. The work of MQ and
LS~is also partly supported by Spanish MINEICO under Grant CICYT-FEDER-FPA2014-55613-P, and by the Severo Ochoa Excellence Program of MINEICO under Grant SO-2012-0234. The research of EM is also partly supported by Spanish MINEICO under Grant FPA2015-64041-C2-1-P, and by the Basque Government under Grant IT979-16.
\end{acknowledgments}


\end{document}